\documentclass[10pt,a4]{article}
\usepackage[dvips]{graphicx,color}
\usepackage{amsmath,bm,amssymb,calc}

\begin{document}

\title{Predicting magic numbers of nuclei
with semi-realistic nucleon-nucleon interactions}


\author{H. Nakada\footnote{E-mail: nakada@c.chiba-u.ac.jp}
and K. Sugiura\\
\textit{Department of Physics, Graduate School of Science and Technology, Chiba University,}\\
\textit{Yayoi-cho 1-33, Inage, Chiba 263-8522, Japan}}


\maketitle

\begin{abstract}%
Magic numbers are predicted in wide range of the nuclear chart
by the self-consistent mean-field calculations
with the M3Y-P6 and P7 semi-realistic $NN$ interactions.
The magic numbers are identified by vanishing pair correlations
in the spherical Hartree-Fock-Bogolyubov regime.
We also identify submagic numbers
when energy gain due to the pairing is sufficiently small.
It is found that the results with M3Y-P6 well correspond to the known data,
apart from a few exceptions.
For some of the magic or submagic numbers the prediction differs
from that with the Gogny-D1S or D1M interaction.
Roles of the tensor force
and the spin-isospin channel originating from the one-pion exchange potential
are investigated in the $Z$- or $N$-dependence of the shell gap.
\end{abstract}

\section{Introduction}\label{sec:intro}

The shell structure is one of the fundamental concepts in nuclear physics,
which is well described as if the constituent nucleons
moved inside the nuclear mean field (MF) almost independently.
A large energy gap between the single-particle (s.p.) orbits
in the spherically symmetric MF produces a magic number,
\textit{i.e.} relative stability of nuclei
having a specific proton ($Z$) or neutron number ($N$),
manifesting the shell structure.
It has been disclosed by the experiments using the radioactive beams
that the shell structure, and therefore the magic numbers,
may depend on $Z$ and $N$~\cite{ref:SP08}.
The magic numbers of nuclei are important
in understanding the origin of matters,
since elements were synthesized via nuclear reactions
which should greatly be influenced by stability of relevant nuclei.
For instance, several peaks in the abundance of elements
correspond to the magic numbers of $N$.
The magic numbers of nuclei could also be responsible
for the existence limit of superheavy elements
that have not yet been observed.

For the $Z$- and $N$-dependence of the shell structure,
which is sometimes called \textit{shell evolution}~\cite{ref:Ots08},
several possible mechanisms have been pointed out.
Since the height of the centrifugal barrier depends
on the orbital angular momentum $\ell$ of the s.p. orbits,
the s.p. energies of the loosely bound orbits
may have significant $\ell$-dependence,
possibly influencing the shell evolution~\cite{ref:N16}.
However, there has been no clear evidence that
the $\ell$-dependence of the s.p. energies gives rise to
appearance or disappearance of magic numbers.
On the other hand,
roles of specific channels in the effective nucleonic interaction
have been argued.
It has been clarified that the tensor channels are important
in the shell evolution~\cite{ref:Vtn,ref:Nak10b,ref:NSM13}.
The central spin-isospin channel of the nucleon-nucleon ($NN$) interaction
has also been considered~\cite{ref:Vst}.
In addition to the $NN$ interaction,
roles of the three-nucleon ($NNN$) interaction
have been investigated~\cite{ref:V3N}.

While significance of the tensor force in the shell evolution
has been clarified qualitatively in Ref.~\cite{ref:Vtn},
effects of specific channels on magic numbers should be assessed carefully,
because in practice magic numbers emerge via interplay of various channels.
One of the authors (H.N.) has developed
the M3Y-type semi-realistic $NN$ interactions
in a series of articles~\cite{ref:Nak03,ref:Nak08b,ref:Nak10,ref:Nak13}.
By applying the numerical methods of Refs.~\cite{ref:NS02,ref:Nak06,ref:Nak08},
the self-consistent Hartree-Fock (HF)
and Hartree-Fock-Bogolyubov (HFB) calculations have been implemented
with the semi-realistic interactions.
Since the interactions explicitly include the tensor force
with realistic origin
and have reasonable nature on the central spin-isospin channel,
the MF approaches employing the semi-realistic interactions
provide us with a suitable framework
for investigating the magic numbers in wide range of the nuclear chart.
A good example is the s.p. level inversion observed from $^{40}$Ca to $^{48}$Ca,
which is reproduced remarkably well by the M3Y-type interactions
including the realistic tensor force~\cite{ref:NSM13}.
Note also that the numerical methods
of Refs.~\cite{ref:NS02,ref:Nak06,ref:Nak08} have ability
to handle loosely bound orbitals.
The MF calculations with the semi-realistic interactions
have been implemented to investigate the shell structure
in neutron-rich Ca and Ni nuclei~\cite{ref:Nak10b}.
In this paper we extensively apply the spherical MF approaches
with the semi-realistic interactions,
particularly the parameter-sets M3Y-P6 and P7~\cite{ref:Nak13},
and predict magic numbers in wide range of the nuclear chart,
from relatively light to heavy nuclei
including nuclei far off the $\beta$ stability.
As discarded in the conventional MF approaches,
tensor-force effects are one of the current hot topics
in nuclear structure physics.
We investigate effects of the tensor force
and those of the spin-isospin channel by comparing the results
to those with the Gogny-D1S~\cite{ref:D1S} and D1M~\cite{ref:D1M} interactions.

\section{Effective Hamiltonian}\label{sec:effH}

Throughout this paper
the Hamiltonian is taken to be $H=H_N+V_C-H_\mathrm{c.m.}$,
consisting of the nuclear Hamiltonian $H_N$,
the Coulomb interaction $V_C$
and the center-of-mass (c.m.) Hamiltonian $H_\mathrm{c.m.}$.
The following non-relativistic form is assumed for $H_N$,
\begin{equation}\label{eq:H_N}
H_N = K + V_N\,;\quad K = \sum_i \frac{\mathbf{p}_i^2}{2M}\,,\quad
V_N = \sum_{i<j} v_{ij}\,,
\end{equation}
and
\begin{equation}\label{eq:effint}
\begin{split}
 v_{ij} &= v_{ij}^{(\mathrm{C})}
 + v_{ij}^{(\mathrm{LS})} + v_{ij}^{(\mathrm{TN})}
 + v_{ij}^{(\mathrm{DD})}\,;\\
v_{ij}^{(\mathrm{C})} &= \sum_n \big(t_n^{(\mathrm{SE})} P_\mathrm{SE}
+ t_n^{(\mathrm{TE})} P_\mathrm{TE} + t_n^{(\mathrm{SO})} P_\mathrm{SO}
+ t_n^{(\mathrm{TO})} P_\mathrm{TO}\big)
 f_n^{(\mathrm{C})} (r_{ij})\,,\\
v_{ij}^{(\mathrm{LS})} &= \sum_n \big(t_n^{(\mathrm{LSE})} P_\mathrm{TE}
 + t_n^{(\mathrm{LSO})} P_\mathrm{TO}\big)
 f_n^{(\mathrm{LS})} (r_{ij})\,\mathbf{L}_{ij}\cdot
(\mathbf{s}_i+\mathbf{s}_j)\,,\\
v_{ij}^{(\mathrm{TN})} &= \sum_n \big(t_n^{(\mathrm{TNE})} P_\mathrm{TE}
 + t_n^{(\mathrm{TNO})} P_\mathrm{TO}\big)
 f_n^{(\mathrm{TN})} (r_{ij})\, r_{ij}^2 S_{ij}\,,\\
v_{ij}^{(\mathrm{DD})} &= \big(t_\rho^{(\mathrm{SE})} P_\mathrm{SE}\cdot
 [\rho(\mathbf{r}_i)]^{\alpha^{(\mathrm{SE})}}
 + t_\rho^{(\mathrm{TE})} P_\mathrm{TE}\cdot
 [\rho(\mathbf{r}_i)]^{\alpha^{(\mathrm{TE})}}\big)
 \,\delta(\mathbf{r}_{ij})\,,
\end{split}
\end{equation}
where $i$ and $j$ are the indices of individual nucleons,
$M=(M_p+M_n)/2$ with $M_p$ ($M_n$) representing
the mass of a proton (a neutron)~\cite{ref:PDG06},
$\mathbf{s}_i$ is the spin operator of the $i$-th nucleon,
$\mathbf{r}_{ij}= \mathbf{r}_i - \mathbf{r}_j$,
$r_{ij}=|\mathbf{r}_{ij}|$,
$\mathbf{p}_{ij}= (\mathbf{p}_i - \mathbf{p}_j)/2$,
$\mathbf{L}_{ij}= \mathbf{r}_{ij}\times \mathbf{p}_{ij}$,
and $\rho(\mathbf{r})$ denotes the nucleon density.
The tensor operator is defined by
$S_{ij}= 4\,[3(\mathbf{s}_i\cdot\hat{\mathbf{r}}_{ij})
(\mathbf{s}_j\cdot\hat{\mathbf{r}}_{ij})
- \mathbf{s}_i\cdot\mathbf{s}_j ]$
with $\hat{\mathbf{r}}_{ij}=\mathbf{r}_{ij}/r_{ij}$.
The $P_\mathrm{SE}$, $P_\mathrm{TE}$, $P_\mathrm{SO}$ and $P_\mathrm{TO}$ operators
indicate the projection on the singlet-even (SE), triplet-even (TE),
singlet-odd (SO) and triplet-odd (TO) two-particle states.
The c.m. Hamiltonian is $H_\mathrm{c.m.}=\mathbf{P}^2/2AM$,
where $\mathbf{P}=\sum_i \mathbf{p}_i$ and $A=Z+N$.
In the M3Y-type interactions
$f_n^{(\mathrm{X})}$ is taken to be the Yukawa function
for all of $\mathrm{X}=\mathrm{C},\,\mathrm{LS},\,\mathrm{TN}$.
In the Gogny-D1S and D1M interactions
we have $f_n^{(\mathrm{C})}(r)=e^{-(\mu_n r)^2}$,
$f^{(\mathrm{LS})}(r)=\nabla^2\delta(\mathbf{r})$ and $v^{(\mathrm{TN})}=0$.
The density-dependent contact term $v^{(\mathrm{DD})}$ carries
certain effects of the $NNN$ interaction
and the nuclear medium (as introduced via the $G$-matrix),
which are significant to reproduce the saturation properties.
It is emphasized that $v^{(\mathrm{TN})}$ is quite realistic in M3Y-P6 and P7,
since it is kept unchanged from the M3Y-Paris interaction~\cite{ref:M3Y-P}
which has the $G$-matrix origin.
The longest-range part in $v^{(\mathrm{C})}$ is also realistic,
maintained to be the corresponding term
of the one-pion exchange potential (OPEP)
and denoted by $v_\mathrm{OPEP}^{(\mathrm{C})}$.
Owing to $v_\mathrm{OPEP}^{(\mathrm{C})}$,
the M3Y-type interactions have spin-isospin properties
consistent with the experimental information~\cite{ref:Nak13}.
Adjusted to the microscopic results
of the pure neutron matter~\cite{ref:FP81,ref:APR98},
M3Y-P6 and P7 give reasonable symmetry energy
up to its density dependence~\cite{ref:Nak13}.

\section{Identification of magic and submagic numbers}\label{sec:iden}

Definition of magic numbers is not necessarily clear.
Experimentally, they have been identified by relative stability;
\textit{e.g.} mass irregularity, kink in the separation energies
and high excitation energies.
From theoretical viewpoints,
the magic nature, which is called ``magicity'',
is linked to quenching of the many-body correlations.
Typically, the spherical HF solution is expected
to give a good approximation for the doubly magic nuclei.
The nuclei having either magic $Z$ or $N$ are usually spherical.
For spherical nuclei,
the pairing among like nucleons provides dominant correlation beyond the HF.
We shall therefore identify magic numbers (and submagic numbers)
by comparing the spherical HF and HFB results for even-even nuclei.

The quadrupole deformation can be another source that breaks magicity.
However, both the pair excitation and the quadrupole deformation
are driven by quenching of the shell gap in most cases,
although strength of the residual interaction
associated with the individual correlation is relevant as well.
Thus the sizable pair correlation within the spherical HFB
could also be a measure of the quadrupole deformation.
Conversely, when we view breaking of magicity via the pairing in this study,
it does not necessarily indicate that the breakdown takes place
due to the pair correlation in actual.
The present work would give informative overview
of distribution of magic numbers over the nuclear chart.
Although more precise investigation by taking into account
the deformation degrees of freedom (d.o.f) will be desirable,
we leave it as a future study
which needs intensive calculations.

The magic numbers $Z=2,8,20,28,50,82$ and $N=2,8,20,28,50,82,126$
have been established around the $\beta$ stability line.
These numbers are kept to be magic in certain region,
whereas it has been clarified by experiments
that the $N=8,20,28$ magicity is eroded far off the $\beta$ stability.
As will be shown in Sec.~\ref{sec:magic},
$N=184$ can also be a good magic number, insensitive to $Z$.
We first implement the HF and HFB calculations
for even-even nuclei along these numbers.
If we find a good candidate for another magic number,
we additionally carry out calculations around it.
Since validity of the MF approaches could be questioned in very light nuclei
and it is difficult to collect experimental data for extremely heavy nuclei,
we restrict ourselves to the nuclei having $8\leq Z\leq 126$ and $N\leq 200$.

From the spherical HFB results, we identify $Z$ ($N$) to be \textit{magic}
when the proton (neutron) pair correlation vanishes.
We recognize the vanishing pair correlation
through the proton (neutron) pair energy
$E_p^\mathrm{pair}$ ($E_n^\mathrm{pair}$),
\textit{i.e.} the energy contributed by the proton (neutron) pairing tensor
in the HFB state.
There are certain cases in which the pair correlation survives,
but the HF and the HFB energies,
which are denoted by $E_\mathrm{HF}$ and $E_\mathrm{HFB}$, are very close.
In these nuclei correlation effects are suppressed,
resulting in \textit{e.g.} relatively high excitation energy.
Such suppression takes place
particularly when either of $Z$ or $N$ is a good magic number,
as has been pointed out for $^{68}$Ni~\cite{ref:Nak10b}
and $^{146}$Gd~\cite{ref:MNOM}.
We therefore identify $Z$ ($N$) to be \textit{submagic}
if $E_\mathrm{HF}-E_\mathrm{HFB}$ is smaller than a certain value
$\lambda_\mathrm{sub}$ for $N=\mbox{magic}$ ($Z=\mbox{magic}$) nuclei.
We adopt $\lambda_\mathrm{sub}=0.5\,\mathrm{MeV}$
and $0.8\,\mathrm{MeV}$ in the calculations below, independent of $A$.
Comparison of the results using different $\lambda_\mathrm{sub}$ values
will exhibit how they are sensitive (or insensitive) to $\lambda_\mathrm{sub}$.

There could be disputes on the above criterion
for the magic and submagic numbers.
It is true that the criterion based solely on the spherical MF calculations
is not complete.
We may find influence of quadrupole deformation
if comparing the current D1S results
to those of the comprehensive deformed HFB calculations~\cite{ref:D1S-Web}.
However, previous studies have suggested~\cite{ref:Nak10b}
that the semi-realistic interactions often give simple picture
for appearance and disappearance of the magicity,
in good connection to the shell structure under the spherical symmetry.
Indeed, prediction of magicity with the semi-realistic interactions
is in good harmony with the known data,
particularly for relatively light-mass region.
Moreover, since quadrupole deformation may quench magicity
but cannot enhance it,
the prediction within the spherical MF regime is useful
in selecting candidates and in overviewing
how the magic numbers can distribute over the nuclear chart.

There were several works in which the magicity was studied
from kinks in the two-proton ($S_{2p}$) or two-neutron ($S_{2n}$)
separation energies~\cite{ref:magic-S2n}.
Whereas $S_{2p}$ and $S_{2n}$ are calculable
within the spherical HFB calculations,
it is not straightforward to draw a quantitative measure for magicity
from $S_{2p}$ and $S_{2n}$
that is applicable to wide range of the nuclear chart.
We here comment that the magic and submagic numbers shown below,
which are identified from the criterion given above,
are compatible with the kinks in the calculated $S_{2p}$ or $S_{2n}$.

\section{Predicting magic and submagic numbers}\label{sec:magic}

We now show prediction on magic and submagic numbers,
which are identified by the criterion given in Sec.~\ref{sec:iden}.
For the HF and HFB calculations,
we have used the methods developed in Ref.~\cite{ref:Nak06}
with the basis functions of Ref.~\cite{ref:Nak08}.
The s.p. bases up to $\ell=\ell_\mathrm{max}+2$,
where $\ell_\mathrm{max}$ is the highest $\ell$ of the occupied level
in the HF configuration,
should be included for each nucleus to handle the pair correlation
appropriately~\cite{ref:Nak06}.
We use the $\ell\leq 7$ bases for the $N<82$ nuclei,
the $\ell\leq 8$ bases for the $82\leq N< 126$ nuclei
and the $\ell\leq 9$ bases for the $N\geq 126$ nuclei in the HFB.
No additional approximations are imposed,
by explicitly treating the exchange
and the pairing terms of $V_C$~\cite{ref:Coul-pair}
as well as the $2$-body term of $H_\mathrm{c.m.}$.

\subsection{Overview of magic and submagic numbers}
\label{subsec:magic_overview}

\begin{figure}[!h]
\centerline{\includegraphics[scale=1.2]{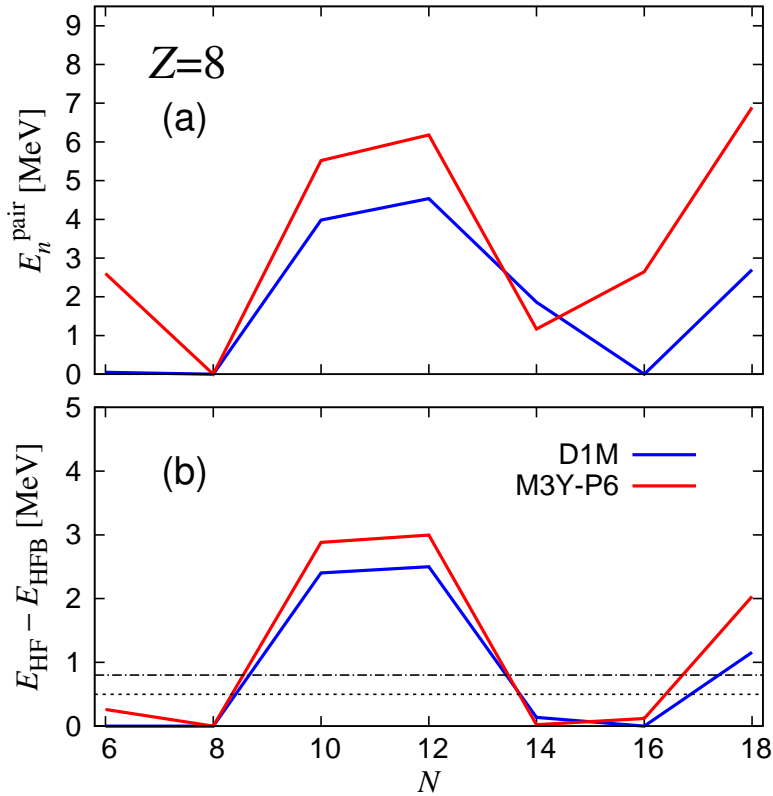}}
\caption{(a) $E_n^\mathrm{pair}$ and (b) $E_\mathrm{HF}-E_\mathrm{HFB}$
 in the $Z=8$ nuclei.
 The blue and red lines present the results
 with the D1M and M3Y-P6 interactions, respectively.
 In (b), the $\lambda_\mathrm{sub}$ values
 ($0.5\,\mathrm{MeV}$ and $0.8\,\mathrm{MeV}$)  are shown
 by the thin dotted and dot-dashed lines.}
\label{fig:Epr_Z8}
\end{figure}

\begin{figure}[!h]
\centerline{\includegraphics[scale=1.2]{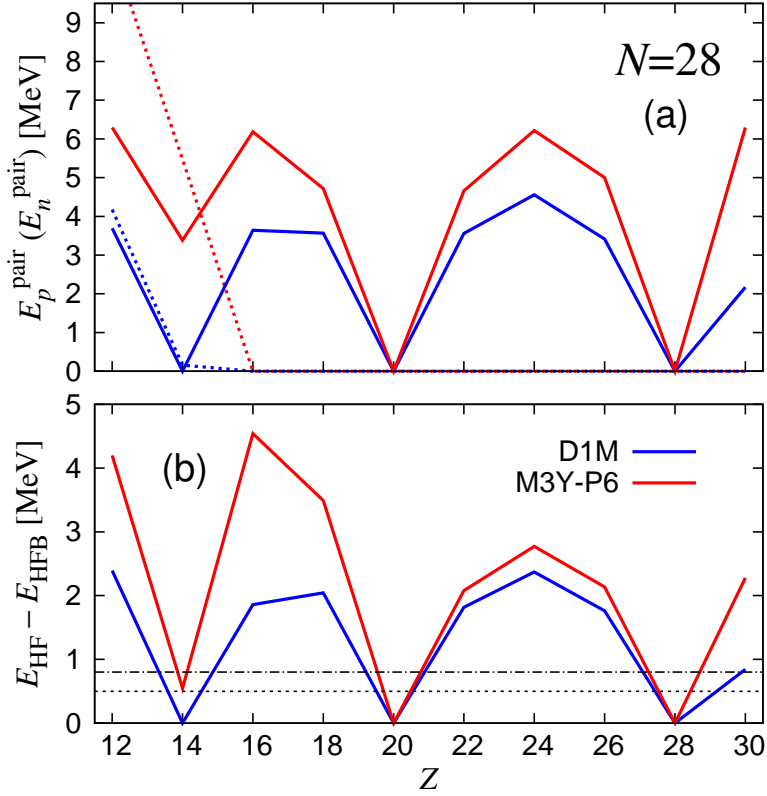}}
\caption{(a) $E_p^\mathrm{pair}$ ($E_n^\mathrm{pair}$)
 and (b) $E_\mathrm{HF}-E_\mathrm{HFB}$ in the $N=28$ nuclei.
 In (a) $E_n^\mathrm{pair}$ is presented by the dotted line
 while $E_p^\mathrm{pair}$ by the solid lines.
 Convention of colors is similar to Fig.~\protect\ref{fig:Epr_Z8}.}
\label{fig:Epr_N28}
\end{figure}

To illustrate how we identify magic and submagic numbers,
we show $E_n^\mathrm{pair}$ and $E_\mathrm{HF}-E_\mathrm{HFB}$
for the O isotopes calculated with D1M and M3Y-P6,
in Fig.~\ref{fig:Epr_Z8}.
Likewise for the $N=28$ isotones in Fig.~\ref{fig:Epr_N28}.

In the O isotopes we always have $E_p^\mathrm{pair}=0$
and therefore $Z=8$ is regarded to be magic.
Because $E_n^\mathrm{pair}$ vanishes at $N=8$,
$^{16}$O is a good doubly-magic nucleus.
We find that $E_n^\mathrm{pair}$ also vanishes at $N=16$
with the D1M interaction,
indicating that $N=16$ is a magic number at $^{24}$O.
With the M3Y-P6 interaction,
$E_n^\mathrm{pair}$ does not fully vanish though relatively small.
The suppression of the neutron pairing is confirmed
in $E_\mathrm{HF}-E_\mathrm{HFB}$,
and we consider that $N=16$ is submagic at $^{24}$O in the M3Y-P6 result.
Analogously, $N=14$ is submagic both with D1M and M3Y-P6.

In Fig.~\ref{fig:Epr_N28},
$E_n^\mathrm{pair}$ as well as $E_p^\mathrm{pair}$ are presented
for the $N=28$ isotones.
We remark that non-vanishing $E_n^\mathrm{pair}$ is obtained
in the $Z\leq 14$ region with the M3Y-P6 interaction.
The same holds with M3Y-P7, though not shown.
This well corresponds to the observed breakdown
of the $N=28$ magicity in this region.
At $^{42}$Si $E_\mathrm{HF}-E_\mathrm{HFB}$ is less than $0.8\,\mathrm{MeV}$.
However, deformation may be driven because neither $Z$ nor $N$ is magic.
For this reason we do not regard $N=28$ to be submagic at $^{42}$Si,
although deformed MF calculations are required for full justification.
In contrast, when we apply D1M,
$Z=14$ is magic and $N=28$ remains submagic at $^{42}$Si,
although the $N=28$ magicity is lost at $^{40}$Mg.

The $N=28$ magicity holds in $Z\geq 16$.
We do not find quenching of the proton pairing in this region,
except at the normal magic numbers $Z=20$ and $28$.

\begin{figure}[!hp]
\centerline{\includegraphics[scale=0.75,angle=90]{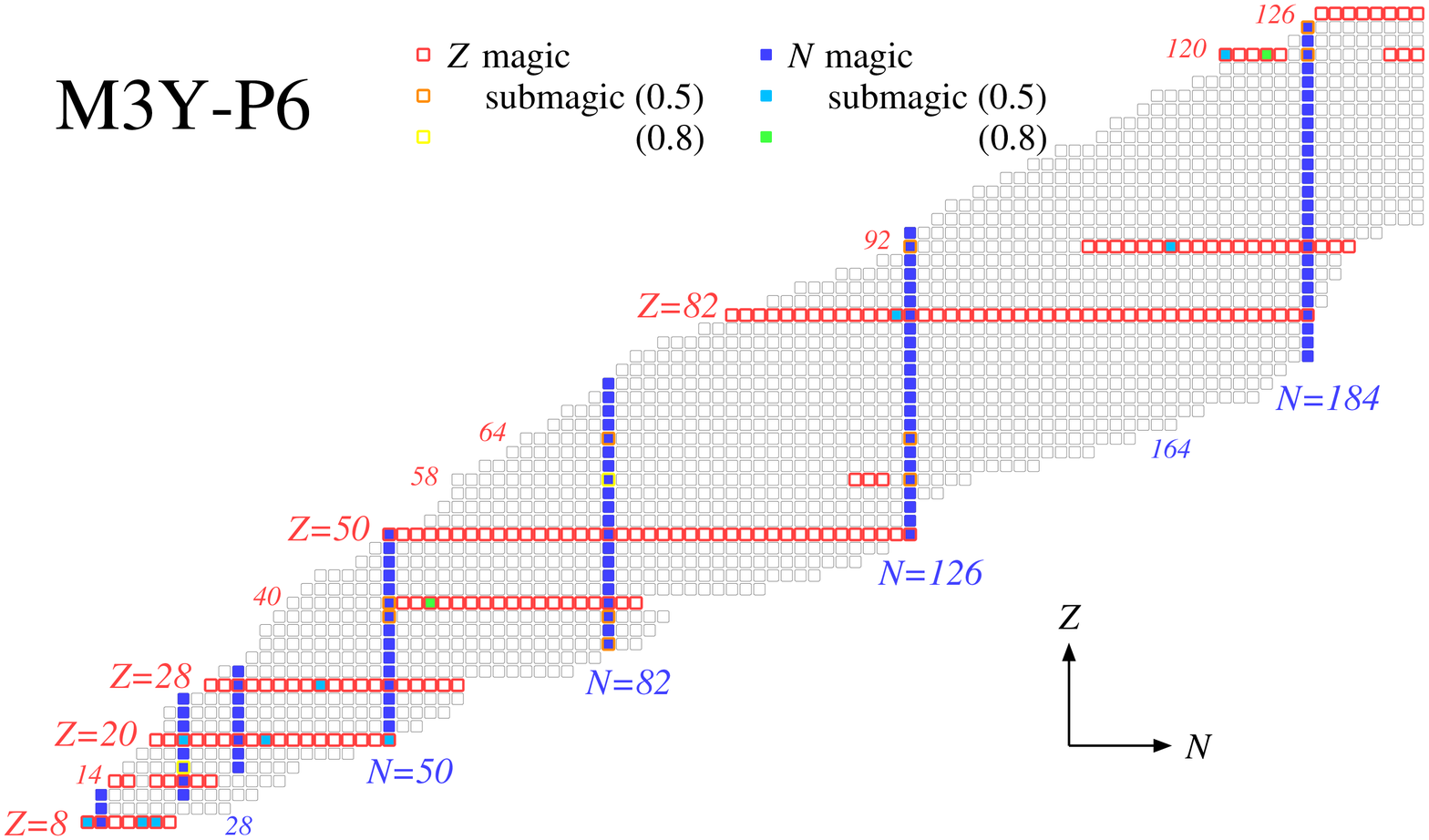}}
\caption{Chart showing magic numbers predicted with the M3Y-P6 interaction.
Individual boxes correspond to even-even nuclei.
Magic (submagic) $Z$'s are represented by the red-colored
(orange- or yellow-colored) frame,
and magic (submagic) $N$'s by filling the box with the blue
(skyblue or green) color.
The $\lambda_\mathrm{sub}$ values for the submagic numbers (in MeV)
are as parenthesized.}
\label{fig:magic_M3Yp6}
\end{figure}

\begin{figure}[!hp]
\centerline{\includegraphics[scale=0.75,angle=90]{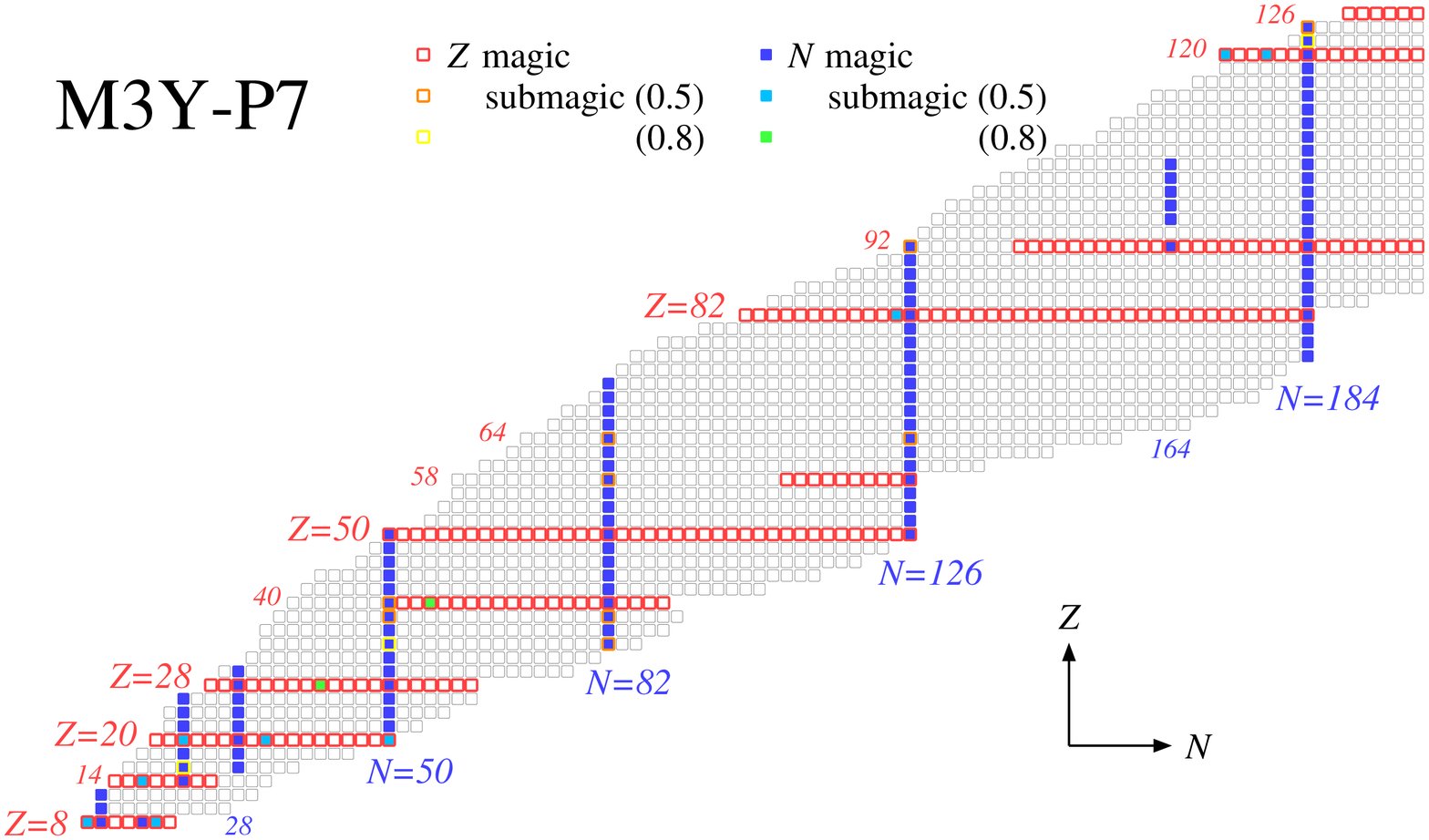}}
\caption{Chart showing magic numbers predicted with the M3Y-P7 interaction.
See Fig.~\protect\ref{fig:magic_M3Yp6} for conventions.}
\label{fig:magic_M3Yp7}
\end{figure}

\begin{figure}[!hp]
\centerline{\includegraphics[scale=0.75,angle=90]{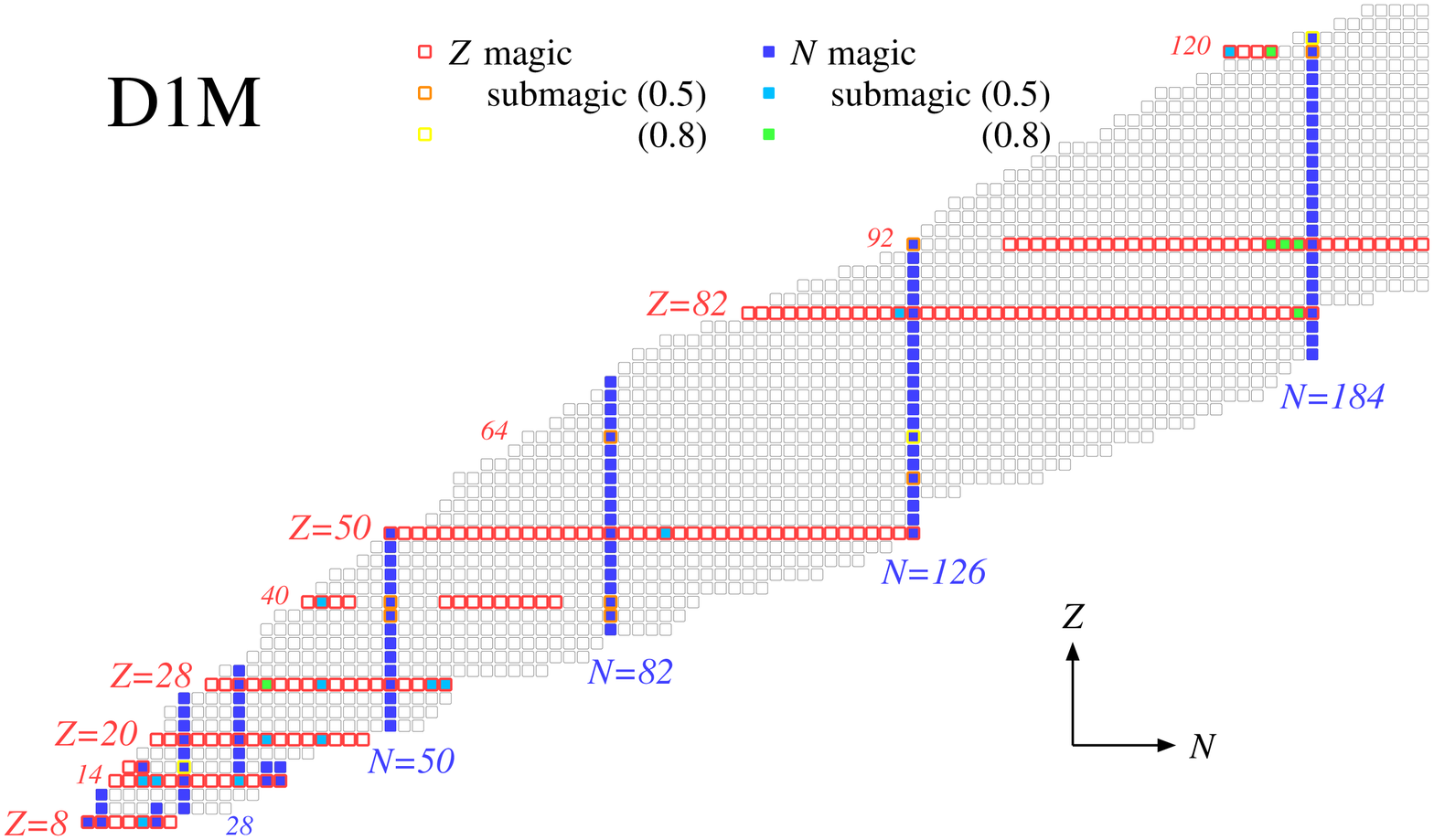}}
\caption{Chart showing magic numbers predicted with the D1M interaction.
See Fig.~\protect\ref{fig:magic_M3Yp6} for conventions.}
\label{fig:magic_D1M}
\end{figure}

We search magic and submagic numbers in this manner
in the wide range of the nuclear chart,
applying the spherical HF and HFB calculations.
The prediction with the M3Y-P6 and P7 semi-realistic interactions
is summarized in Figs.~\ref{fig:magic_M3Yp6} and \ref{fig:magic_M3Yp7}.
For comparison, the prediction with D1M is displayed
in Fig.~\ref{fig:magic_D1M}.
We shall take a close look at these results,
along with the result with D1S, in the subsequent subsections.
The boundaries in Figs.~\ref{fig:magic_M3Yp6}\,--\,\ref{fig:magic_D1M}
for the nuclei having magic $Z$ or $N$
are obtained from the spherical HFB calculations.
For open-shell nuclei the boundaries are somewhat arbitrary,
not representing reliable drip lines,
because we have not taken into account deformation in the present work.

\subsection{$N=6$, $14$ and $16$}\label{subsec:magic_N6&N14&N16}

The $N=16$ magicity at $^{24}$O observed in the experiments~\cite{ref:N16}
is more or less reproduced by any of the four interactions;
magic in the Gogny interactions (D1S and D1M)
while submagic in the M3Y-P6 and P7 interactions.
The $N=16$ magicity is extended to $Z=10$ with D1M
and even to $Z=14$ with D1S.
We find that $N=14$ as well as $N=6$ behave magic or submagic
in the O nuclei, as seen in Fig.~\ref{fig:Epr_Z8}.

\subsection{$N=20$ and $28$}\label{subsec:magic_N20&N28}

It is remarked that the $N=20$ magicity is lost
at $Z=10$ with M3Y-P6 and at $Z=10,12$ with M3Y-P7,
though not with D1M.
This reminds us of the experimentally established `island of inversion'.
Usual interpretation of the island of inversion
has been the deformation around $N=20$
for the $Z\lesssim 12$ nuclei~\cite{ref:N20-def},
whereas possibility of the large fluctuation of the pairing
has also been argued~\cite{ref:Mg32-sph}.
Although the present calculation does not tell us its mechanism,
it is interesting to view that the loss of the magicity
is well described with M3Y-P7, and partially with M3Y-P6.

Another notable point is that $N=20$ becomes submagic at $^{40}$Ca,
rather than magic, with M3Y-P6 and M3Y-P7.
This point should further be investigated in the near future.

As already argued in relation to Fig.~\ref{fig:Epr_N28},
the $N=28$ magicity is lost in $N\leq 14$ in the M3Y-P6 and P7 results,
while only in $N\leq 12$ in the D1S and D1M results,
although quadrupole deformation has been shown to emerge at $^{42}$Si
in the deformed HFB calculation with D1S~\cite{ref:D1S-Web}.

\subsection{$N=32$, $34$ and $40$}\label{subsec:magic_N32&N34&N40}

Although $N=32$ and $34$ are magic for the Si and the S isotopes
with D1M (so is $N=34$ with D1S),
this prediction is not supported
with the semi-realistic M3Y-P6 and P7 interactions.
Irrespective of the interactions, $N=32$ is submagic at $^{52}$Ca
owing to the closure of the $n1p_{3/2}$ orbit,
as is consistent with the experimental data~\cite{ref:Ca52_Ex2}.
However, the $N=32$ magicity is not found at $^{60}$Ni with M3Y-P6 and P7,
though kept with D1S and D1M.
Origin of the $Z$- and the interaction-dependence of the $N=32$ magicity
will be discussed in Sec.~\ref{sec:TN-OPEP}.
The recently suggested magicity of $N=34$ at $^{54}$Ca~\cite{ref:Ca54_Ex2}
is not seen in Figs.~\ref{fig:magic_M3Yp6}\,--\,\ref{fig:magic_D1M}.

The submagic nature of $N=40$ at $^{68}$Ni,
which is again compatible with the data~\cite{ref:Ni68},
is reproduced by all the interactions.
On the other hand, the $N=40$ magicity is lost at $^{60}$Ca
in the M3Y-P6 and P7 results,
while it is preserved in the D1S and D1M results.
The $N=40$ magicity depends on the interactions also at $^{80}$Zr,
to which we shall return in Sec.~\ref{subsec:magic_Z34&Z38&Z40}.

\subsection{$N=50$, $56$ and $58$}\label{subsec:magic_N50&N56&N58}

We do not find loss of the $N=50$ magicity except at $^{70}$Ca.
It is noted that $N=50$ is indicated to be submagic at $^{70}$Ca,
rather than magic, with the M3Y-P6 and P7 interactions.
The $^{70}$Ca nucleus is not bound within the spherical HFB with D1S and D1M.

We find that $N=56$ is submagic at $^{96}$Zr except with D1M
because of the $n1d_{5/2}$ closure,
well corresponding to the high first excitation energy
in the measurements~\cite{ref:NDS96}.
In the D1S and D1M results $N=56$ and $58$ are submagic in the Ni isotopes.
Although the submagic nature of $N=58$ has been argued using M3Y-P5
in Ref.~\cite{ref:Nak10b},
the magicity is not apparent in the M3Y-P6 and P7 results.

\subsection{$N=82$, $90$, $124$ and $126$}
\label{subsec:magic_N82&N90&N124&N126}

The magicity of $N=82$ and $N=126$ is maintained in the whole region of $Z$.

$N=90$ is predicted to be submagic with D1S and D1M at $^{140}$Sn
because the $n1f_{7/2}$ orbit is fully occupied,
but not with M3Y-P6 and P7.
$N=124$ becomes submagic at $^{206}$Pb with all the interactions,
having the $(n2p_{1/2})^{-2}$ configuration.

\subsection{$N=164$, $184$ and in-between}\label{subsec:magic_N164&N184}

$N=164$ becomes magic with M3Y-P7 at $^{256}$U and submagic with M3Y-P6
because of the occupation of the $n0i_{11/2}$, $n1g_{9/2}$
and $n0j_{15/2}$ orbits,
while it does not so with the D1S and D1M interactions.
The $N=164$ magicity is developed in the M3Y-P7 results,
occurring also in $96\leq N\leq 104$.
The neutron numbers $N=178$, $180$ and $182$
are classified to be submagic for the U nuclei with D1M,
all of which are not with the M3Y-P6 and P7 interactions.
The $N=178$ magicity is connected to the occupation
of $n1g_{7/2}$ and $n2d_{5/2}$,
and $N=182$ is to the additional occupation of $n2d_{3/2}$,
leaving $n3s_{1/2}$ unoccupied.
However, $N=180$ is not connected to the subshell closure.
The pair correlation seems to come small
partly because of the large mass number and the relatively small degeneracy
of the relevant orbitals.
This problem in the D1M result could be lifted
if we introduce $A$-dependence in $\lambda_\mathrm{sub}$,
though it is not needed for M3Y-P6.

Along $Z=120$, $N=172$ and $178$ become submagic with all the interactions.
With D1S, $^{292}120$ is even a doubly magic nucleus
and $N=182$ is also submagic.

$N=184$ is a good magic number for all the bound nuclei,
with any of the interactions employed in the present work,
whereas the deformed HFB calculations with D1S have suggested
instability against fission in $Z\geq 104$~\cite{ref:D1S-Web}.

\subsection{$Z=14$ and $16$}\label{subsec:magic_Z14&Z16}

Having $Z=14$ and $N=20$, $^{34}$Si is predicted to be doubly magic
with any of the four interactions,
as has been argued in connection
to the proton bubble structure~\cite{ref:NSM13}.
$Z=14$ stays as a magic number in the results
with the D1S and D1M interactions.
In contrast, it is not a magic number in $N\geq 26$ with M3Y-P6 and P7,
nor at $N=14$ with M3Y-P6.
The stiffness of the $Z=14$ core seems relevant
to where the $N=28$ magicity is broken, which has been argued
in Sec.~\ref{subsec:magic_N20&N28}.

All the interactions indicate that $Z=16$ is weakly submagic at $^{36}$S.
With the D1M interaction $Z=16$ becomes magic at $N=12$ and $14$
(also at $N=16$ with D1S),
while it is not with the M3Y-P6 and P7 interactions.

\subsection{$Z=20$ and $28$}\label{subsec:magic_Z20&Z28}

In the present study,
$Z=20$ and $28$ remain to be magic numbers in the whole region of $N$
and irrespective of the effective interactions.
As pointed out in Ref.~\cite{ref:Nak10b},
the persistence of the $Z=28$ magicity around $^{78}$Ni
is contrasted to the argument in Ref.~\cite{ref:Vtn},
although the realistic tensor force is included in M3Y-P6 and P7.
This difference happens because magic numbers are a result
of interplay of various interaction channels,
even though the tensor force plays a significant role.

\subsection{$Z=34$, $38$ and $40$}\label{subsec:magic_Z34&Z38&Z40}

The $Z=38$ and $40$ magicity has been known to be enhanced
along the $N=50$ isotones.
This nature is well described, with both $Z$'s staying submagic.

We find that $Z=34$ is submagic with M3Y-P6 and P7 at $N=82$
while $^{116}$Se is unbound with D1S and D1M.
$Z=38$ is predicted to be submagic with all of the four interactions
at $^{120}$Sr.

The Zr isotopes have been known to exhibit
remarkable $N$-dependence in their structure.
While $^{90}$Zr is close to doubly magic,
the Zr nuclei are deformed in $60\leq N\lesssim 70$
as well as in $N\approx 40$.
We find that the current results for the Zr isotopes
significantly depend on the interactions.
Although $Z=40$ is magic in $38\leq N\leq 46$ with D1M
(in $38\leq N\leq 48$ with D1S),
this magicity is broken with M3Y-P6 and P7,
which is consistent with the experimental data.
As a typical example,
$^{80}$Zr is doubly magic with the Gogny interactions,
while both $Z$ and $N$ lose magicity with the M3Y-P6 and P7 interactions.
The observed energy levels show
that $^{80}$Zr is deformed and never doubly magic.
The deformed HFB does not solve this problem
of the D1S interaction~\cite{ref:D1S-Web}.

In the present work using the spherical MF calculations,
$Z=40$ is indicated to be magic in $60\lesssim N\lesssim 70$,
with any of the four interactions.
This is contradictory to the recent experiments~\cite{ref:Zr106-108}.
It is desired to implement deformed MF calculations,
particularly those using the semi-realistic interactions,
to check whether and how the experimental data is accounted for.

The $^{122}$Zr nucleus is doubly magic in the M3Y-P6 and P7 results,
and is close to doubly magic with submagic nature of $Z=40$
in the Gogny results.

\subsection{$Z=50$, $58$ and $64$}\label{subsec:magic_Z50&Z58&Z64}

We predict no breakdown of the $Z=50$ magicity in the present calculations.
However, deformation has been suggested for neutron-rich Sn nuclei
in the previous calculations; \textit{e.g.} for $98\leq N\leq 110$
in the calculations with D1S~\cite{ref:D1S-Web}
and similarly in the relativistic MF calculations~\cite{ref:GTM04}.
Further study is desirable by applying the M3Y-P6 and P7 interactions
to the deformed MF calculations.

All the current interactions reproduce
the submagic nature of $Z=64$ at $^{146}$Gd,
which has long been investigated (\textit{e.g.} Ref.~\cite{ref:MNOM}).
We also find that $Z=58$ becomes submagic at $^{140}$Ce
with M3Y-P6 and P7, though not with D1S and D1M.
The measured first excitation energy is slightly higher
in $^{140}$Ce~\cite{ref:NDS140}
than in the surrounding $N=82$ isotones $^{138}$Ba and $^{142}$Nd.

Irrespective of the interactions,
we view certain magicity of $Z=58$ at $^{184}$Ce
(magic with M3Y-P7 and submagic with the others),
and of $Z=64$ at $^{190}$Gd (submagic with any of the interactions).
$Z=58$ becomes magic in $118\leq N\leq 122$ with M3Y-P6
and in $108\leq N\leq 126$ with M3Y-P7,
though not in the D1S and D1M results.

\subsection{$Z=82$ and $92$}\label{subsec:magic_Z82&Z92}

Within the spherical MF calculations,
the Pb nuclei are bound up to $N=184$ with all the interactions,
and the last bound nucleus $^{266}$Pb is predicted
to be a good doubly magic nucleus.
Since quadrupole deformation has been predicted to take place
in $144\leq N\leq 166$ by the deformed HFB calculations
with D1S~\cite{ref:D1S-Web},
stability against deformation should further be investigated in future studies.

The $Z=92$ number gains certain magicity,
behaves as submagic at $^{218}$U.
Whereas $Z=92$ is not fully closed in the neutron-deficient region,
it is predicted to be magic in $N\gtrsim 150$ in the present work.
This magicity is stronger in the D1S, D1M and M3Y-P7 results,
even holding at $^{238}$U
which is a well-known deformed nucleus~\cite{ref:NDS238},
in contrast to the M3Y-P6 result in which $Z=92$ is not magic up to $N=150$.
It should be mentioned that deformed ground states have been predicted
by the deformed HFB calculations with D1S in $N\lesssim 170$~\cite{ref:D1S-Web}.

\subsection{$Z=120$, $124$ and $126$}\label{subsec:magic_Z120&Z124&Z126}

The proton magic numbers beyond $Z=100$ have attracted interest,
in connection to the superheavy nuclei
in the so-called `island of stability'.
Although $Z=114$ has been considered a candidate of a magic number,
the present work does not support it
irrespectively of the interactions.
On the other hand, $Z=120$ may behave as a magic number.
We view that $Z=120$ is magic in $N\leq 178$
with any of the four interactions.
In the M3Y-P7 result the $Z=120$ magicity extends to $N=200$,
and in the M3Y-P6 result it disappears at $N=180$
but revives at $N=196$, apart from its the submagic nature at $N=184$.
The $Z=120$ magicity is also predicted
in $N\leq 182$ and $N\geq 194$ with D1S.
In the D1M result the $Z=120$ magicity is lost in $N\geq 180$,
despite its submagic nature at $N=184$.

$Z=124$ is predicted to be submagic at $N=184$ with D1S, M3Y-P6 and P7,
while unbound within the spherical HFB with D1M.
$Z=126$ is a good magic number with D1S, M3Y-P6 and M3Y-P7,
but not with D1M.
It is commented that the fission d.o.f. may enter
in the $Z\gtrsim 120$ nuclei~\cite{ref:D1S-Web}.

\section{Effects of tensor and OPEP-central channels}\label{sec:TN-OPEP}

As mentioned in Sec.~\ref{sec:intro},
the tensor force $v^{(\mathrm{TN})}$ in the $NN$ interaction
affects the shell structure to significant degree.
The central spin-isospin channel,
whose dominant part is $v_\mathrm{OPEP}^{(\mathrm{C})}$,
may cooperatively contribute to the shell evolution
in certain cases~\cite{ref:Nak08b}.
In this section we investigate roles of $v^{(\mathrm{TN})}$
and $v_\mathrm{OPEP}^{(\mathrm{C})}$ in some detail,
by analyzing the spherical HF results.
Since these channels are explicitly contained in M3Y-P6 and P7
while not in D1S and D1M,
comparison among these results will be useful
in assessing effects of $v^{(\mathrm{TN})}$ and $v_\mathrm{OPEP}^{(\mathrm{C})}$
on the magicity.

We here comment on the difference in the magicity between M3Y-P6 and P7,
which is found mainly in heavier mass region
in Figs.~\ref{fig:magic_M3Yp6} and \ref{fig:magic_M3Yp7}.
As the $v^{(\mathrm{TN})}$ and $v_\mathrm{OPEP}^{(\mathrm{C})}$ channels
are identical between them,
it is not obvious what is the main source of the difference
between M3Y-P6 and P7.
Although these two parameter-sets yield different neutron-matter energies
at high densities,
it is not likely that this significantly influences the magicity of nuclei.
The strength of $v^{(\mathrm{LS})}$,
which is slightly stronger in M3Y-P7 than in M3Y-P6,
does not account for all the visible difference
between Figs.~\ref{fig:magic_M3Yp6} and \ref{fig:magic_M3Yp7}.

The s.p. energy of the orbit $j$, $\varepsilon_{\tau_z}(j)$ ($\tau_z=p,n$),
is defined by the derivative of the total energy
with respect to the occupation probability.
We extract contribution of $v^{(\mathrm{TN})}$ to $\varepsilon_{\tau_z}(j)$
from the full HF result by
\begin{equation}
 \varepsilon^{(\mathrm{TN})}_{\tau_z}(j)
 = \sum_{j'm'} n_{j'}\langle jmj'm'|v^{(\mathrm{TN})}|jmj'm'\rangle
 = \frac{1}{2j+1}\sum_{j'J} n_{j'}(2J+1)\langle jj'J|v^{(\mathrm{TN})}|jj'J\rangle\,,
\end{equation}
where $n_j$ denotes the occupation probability on $j$.
Likewise for $\varepsilon^{(\mathrm{OPEP})}_{\tau_z}(j)$,
\textit{i.e.} contribution of $v_\mathrm{OPEP}^{(\mathrm{C})}$.
For the shell gap $\mathit{\Delta}\varepsilon_{\tau_z}(j_2\,\mbox{-}\,j_1)
= \varepsilon_{\tau_z}(j_2)-\varepsilon_{\tau_z}(j_1)$,
corresponding quantities
$\mathit{\Delta}\varepsilon^{(\mathrm{TN})}_{\tau_z}(j_2\,\mbox{-}\,j_1)$ and 
$\mathit{\Delta}\varepsilon^{(\mathrm{OPEP})}_{\tau_z}(j_2\,\mbox{-}\,j_1)$
can be considered.
Since the s.p. energies are more or less fitted to the data
in each effective interaction,
the absolute values of
$\mathit{\Delta}\varepsilon^{(\mathrm{TN})}_{\tau_z}(j_2\,\mbox{-}\,j_1)$ or 
$\mathit{\Delta}\varepsilon^{(\mathrm{OPEP})}_{\tau_z}(j_2\,\mbox{-}\,j_1)$
are not very meaningful.
However, $Z$- or $N$-dependence of
$\mathit{\Delta}\varepsilon^{(\mathrm{TN})}_{\tau_z}(j_2\,\mbox{-}\,j_1)$ and
$\mathit{\Delta}\varepsilon^{(\mathrm{OPEP})}_{\tau_z}(j_2\,\mbox{-}\,j_1)$
is important,
which may give rise to the $Z$- or $N$-dependence of the shell gap.
We shall argue several typical cases in which $v^{(\mathrm{TN})}$
and/or $v_\mathrm{OPEP}^{(\mathrm{C})}$ play a significant role
in $Z$- or $N$-dependence of the magicity.

\subsection{$N=16$, $32$ and $40$}\label{subsec:TN-OPEP_N16&N32&N40}

While $N=16$ behaves as magic or submagic at $^{24}$O
irrespectively of the effective interactions,
its magicity depends on the interactions for larger $Z$,
\textit{i.e.} near the $\beta$ stability.
The shell gap $\mathit{\Delta}\varepsilon_n(0d_{3/2}-1s_{1/2})$
is relevant to the $N=16$ magicity.
In Ref.~\cite{ref:Nak08b}
we have shown that $v^{(\mathrm{TN})}$ and $v_\mathrm{OPEP}^{(\mathrm{C})}$
give the $Z$-dependence by using the older parameter-set M3Y-P5.
Analogous results are obtained with the present parameters M3Y-P6 and P7,
which we do not repeat here.

The $N=32$ magicity is basically determined by
$\mathit{\Delta}\varepsilon_n(0f_{5/2}\,\mbox{-}\,1p_{3/2})$
or $\mathit{\Delta}\varepsilon_n(1p_{1/2}\,\mbox{-}\,1p_{3/2})$.
As mentioned in Sec.~\ref{sec:magic}, $N=32$ is submagic at $^{52}$Ca
but not at $^{60}$Ni with the M3Y-P6 and P7 interactions.
We have found that $N=40$ is submagic at $^{68}$Ni but not at $^{60}$Ca
in the M3Y results,
though submagic at both nuclei in the Gogny results.
The highest occupied neutron orbit is $1p_{1/2}$ at $^{68}$Ni
while $0f_{5/2}$ at $^{60}$Ca.
The strong $Z$-dependence of
$\mathit{\Delta}\varepsilon_n(0g_{9/2}\,\mbox{-}\,0f_{5/2})$
due to $v^{(\mathrm{TN})}$
is important to the erosion of the $N=40$ magicity at $^{60}$Ca.
Role of $v^{(\mathrm{TN})}$ and $v_\mathrm{OPEP}^{(\mathrm{C})}$
in the $N=32$ and $40$ magicity
was already presented in Ref.~\cite{ref:Nak10b} by employing M3Y-P5.
Essential points do not change with M3Y-P6 and P7.

\subsection{$N=56$}\label{subsec:TN-OPEP_N56}

\begin{figure}[!h]
\centerline{\includegraphics[scale=1.5]{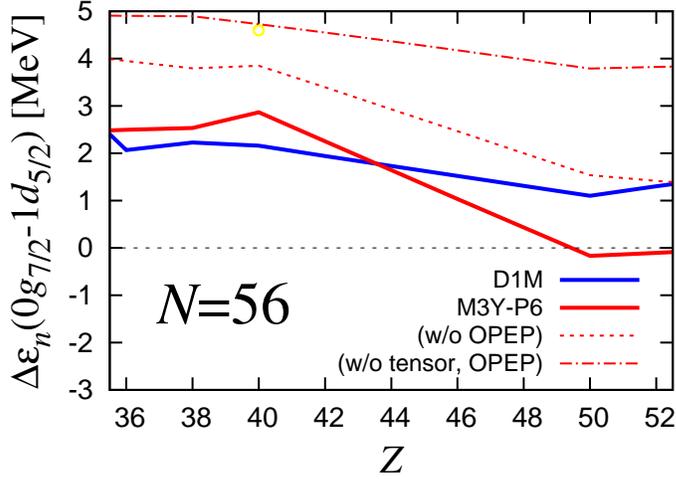}}
\caption{$\mathit{\Delta}\varepsilon_n(0g_{7/2}\,\mbox{-}\,1d_{5/2})$
 in the $N=56$ nuclei.
 Blue and red solid lines are the results with D1M and M3Y-P6, respectively.
 The yellow circle at the top part of the figure indicates
 that $N=56$ is submagic at $Z=40$ in Fig.~\protect\ref{fig:magic_M3Yp6}
 (\textit{i.e.} the M3Y-P6 result).
 Thin red dotted and dot-dashed lines are obtained by subtracting
 contributions of $v^{(\mathrm{TN})}$ and $v_\mathrm{OPEP}^{(\mathrm{C})}$
 from the M3Y-P6 result;
 see text for details.}
\label{fig:den_N56}
\end{figure}

From Figs.~\ref{fig:magic_M3Yp6}\,--\,\ref{fig:magic_D1M}
we have found that the submagic nature of $N=56$ at $^{96}$Zr
is well accounted for with M3Y-P6 and P7, but not with D1S and D1M.
We present the relevant s.p. energy difference
$\mathit{\Delta}\varepsilon_n(0g_{7/2}\,\mbox{-}\,1d_{5/2})$
in Fig.~\ref{fig:den_N56}, for D1M and M3Y-P6.
It is found that $\mathit{\Delta}\varepsilon_n$ strongly depends on $Z$
in the M3Y-P6 result,
as $p0g_{9/2}$ is occupied in $40\leq Z\leq 50$.
This $Z$-dependence produces the relatively large gap at $^{96}$Zr.
The M3Y-P6 result also suggests that $n1d_{5/2}$ and $n0g_{7/2}$
are nearly degenerate around $^{106}$Sn,
compatible with the observed levels in $^{105,107}$Sn~\cite{ref:NDS105&107}.

In order to clarify effects of $v_\mathrm{OPEP}^{(\mathrm{C})}$ and $v^{(\mathrm{TN})}$
on the $Z$-dependence of the shell gap,
$\mathit{\Delta}\varepsilon_n-\mathit{\Delta}\varepsilon_n^{(\mathrm{OPEP})}$
is plotted by the thin red dotted line
and $\mathit{\Delta}\varepsilon_n-\mathit{\Delta}\varepsilon_n^{(\mathrm{OPEP})}
-\mathit{\Delta}\varepsilon_n^{(\mathrm{TN})}$
by the thin red dot-dashed line in Fig.~\ref{fig:den_N56}.
The difference between the red solid line and the red dotted line corresponds
to $\mathit{\Delta}\varepsilon_n^{(\mathrm{OPEP})}(0g_{7/2}\,\mbox{-}\,1d_{5/2})$,
representing effects of $v_\mathrm{OPEP}^{(\mathrm{C})}$,
and the difference between the red dotted line and the red dot-dashed line
to $\mathit{\Delta}\varepsilon_n^{(\mathrm{TN})}(0g_{7/2}\,\mbox{-}\,1d_{5/2})$,
showing effects of $v^{(\mathrm{TN})}$.
It is found that the difference in the s.p. energies depends on $Z$ only weakly
as long as we do not have contributions
of $v_\mathrm{OPEP}^{(\mathrm{C})}$ and $v^{(\mathrm{TN})}$,
both from the D1M result and the M3Y-P6 result
of $\mathit{\Delta}\varepsilon_n-\mathit{\Delta}\varepsilon_n^{(\mathrm{OPEP})}
-\mathit{\Delta}\varepsilon_n^{(\mathrm{TN})}$.
It is reasonable to conclude that
$v^{(\mathrm{TN})}$ and $v_\mathrm{OPEP}^{(\mathrm{C})}$ are responsible
for the $Z$-dependence.

\subsection{$N=164$}\label{subsec:TN-OPEP_N164}

\begin{figure}[!h]
\centerline{\includegraphics[scale=1.5]{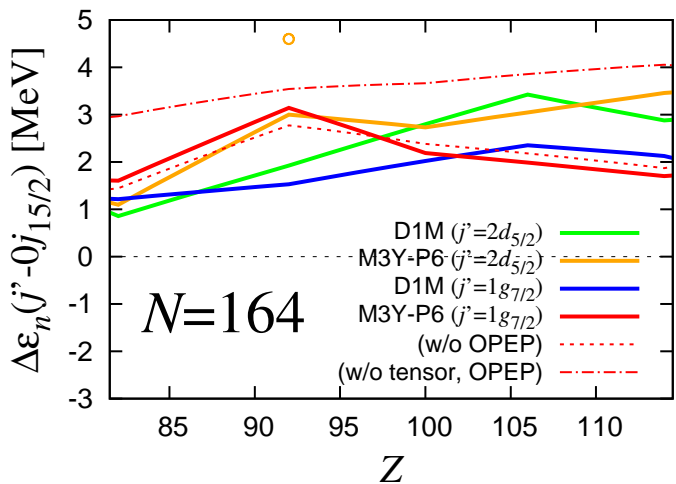}}
\caption{$\mathit{\Delta}\varepsilon_n(j'\,\mbox{-}\,0j_{15/2})$
 ($j'=2d_{5/2}$ or $1g_{7/2}$) in the $N=164$ nuclei.
 Green (orange) solid line is for $j'=2d_{5/2}$ calculated with D1M (M3Y-P6),
 and blue (red) solid line is for $j'=1g_{7/2}$ with D1M (M3Y-P6).
 $N=164$ is submagic at $Z=92$ in Fig.~\protect\ref{fig:magic_M3Yp6},
 which is indicated by the orange circle at the top part of the figure.
 We have no other magicity in Figs.~\protect\ref{fig:magic_M3Yp6}
 and \protect\ref{fig:magic_D1M} (\textit{i.e.} the D1M result).
 Thin red dotted and dot-dashed lines are obtained by subtracting
 contributions of $v^{(\mathrm{TN})}$ and $v_\mathrm{OPEP}^{(\mathrm{C})}$
 in $\mathit{\Delta}\varepsilon_n(1g_{7/2}\,\mbox{-}\,0j_{15/2})$ with M3Y-P6,
 as in Fig.~\protect\ref{fig:den_N56}.}
\label{fig:den_N164}
\end{figure}

As shown in Figs.~\ref{fig:magic_M3Yp6}\,--\,\ref{fig:magic_D1M}
$N=164$ becomes a submagic with M3Y-P6 and a magic number with M3Y-P7
at $^{256}$U,
although it is not with the D1S and D1M interactions.
The $N=164$ shell gap is primarily determined
by $\mathit{\Delta}\varepsilon_n(1g_{7/2}\,\mbox{-}\,0j_{15/2})$,
and is relevant also to
$\mathit{\Delta}\varepsilon_n(2d_{5/2}\,\mbox{-}\,0j_{15/2})$.
These s.p. energy differences are shown for D1M and M3Y-P6
in Fig.~\ref{fig:den_N164}.
We find enhancement of the gap at $Z=92$ in the M3Y-P6 case,
which is regarded as origin of the magicity.
We do not have similar $Z$-dependence in the D1M result.

For $\mathit{\Delta}\varepsilon_n(1g_{7/2}\,\mbox{-}\,0j_{15/2})$,
the M3Y-P6 result of
$\mathit{\Delta}\varepsilon_n-\mathit{\Delta}\varepsilon_n^{(\mathrm{OPEP})}$
(the thin red dotted line)
and $\mathit{\Delta}\varepsilon_n-\mathit{\Delta}\varepsilon_n^{(\mathrm{OPEP})}
-\mathit{\Delta}\varepsilon_n^{(\mathrm{TN})}$
(the thin red dot-dashed line) is plotted in Fig.~\ref{fig:den_N164}.
We find that $\mathit{\Delta}\varepsilon_n(1g_{7/2}\,\mbox{-}\,0j_{15/2})$
is insensitive to $Z$ in the D1M result and in the M3Y-P6 result
after subtracting the contributions of
$v_\mathrm{OPEP}^{(\mathrm{C})}$ and $v^{(\mathrm{TN})}$.
It is thus clarified that the $N=164$ magicity emerges
primarily by $v^{(\mathrm{TN})}$ and complementarily by $v_\mathrm{OPEP}^{(\mathrm{C})}$.

\subsection{$Z=14$}\label{subsec:TN-OPEP_Z14}

\begin{figure}[!h]
\centerline{\includegraphics[scale=1.5]{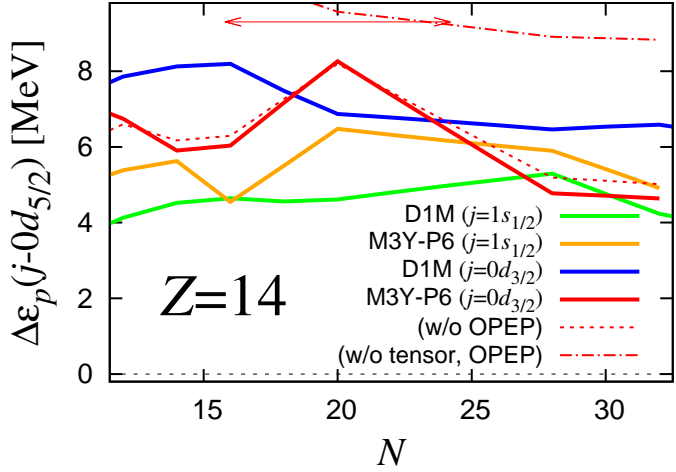}}
\caption{$\mathit{\Delta}\varepsilon_p(j\,\mbox{-}\,0d_{5/2})$
 ($j=1s_{1/2}$ or $0d_{3/2}$) in the $Z=14$ nuclei.
 Green (orange) solid line is for $j=1s_{1/2}$ calculated with D1M (M3Y-P6),
 and blue (red) solid line is for $j=0d_{3/2}$ with D1M (M3Y-P6).
 The region where $Z=14$ is magic in Fig.~\protect\ref{fig:magic_M3Yp6}
 is shown by the red arrow at the top part of the figure,
 while in Fig.~\protect\ref{fig:magic_D1M} $Z=14$ is magic for all $N$.
 $\mathit{\Delta}\varepsilon_p(0d_{3/2}\,\mbox{-}\,0d_{5/2})$'s
 after subtracting the $v^{(\mathrm{TN})}$
 and $v_\mathrm{OPEP}^{(\mathrm{C})}$ contributions from the M3Y-P6 result
 are displayed by thin red dotted and dot-dashed lines,
 as in Fig.~\protect\ref{fig:den_N56}.}
\label{fig:dep_Z14}
\end{figure}

We have shown in Figs.~\ref{fig:magic_M3Yp6}\,--\,\ref{fig:magic_D1M}
that $Z=14$ behaves as a good magic number for all $N$ with D1M,
while its magicity depends on $N$ with M3Y-P6 and P7.
The relevant s.p. energy differences
$\mathit{\Delta}\varepsilon_p(1s_{1/2}\,\mbox{-}\,0d_{5/2})$
and $\mathit{\Delta}\varepsilon_p(0d_{3/2}\,\mbox{-}\,0d_{5/2})$
are presented in Fig.~\ref{fig:dep_Z14} for D1M and M3Y-P6.
It is noted that, while $n0d_{3/2}$ is higher than $n1s_{1/2}$ for all $N$
in the D1M result,
$n0d_{3/2}$ comes down so that the level sequence could be inverted
in $N\geq 26$ in the M3Y-P6 result.
The breakdown of the $Z=14$ magicity at $N\approx 28$ with M3Y-P6
seems linked to this behavior of $n0d_{3/2}$.

Effects of $v_\mathrm{OPEP}^{(\mathrm{C})}$ and $v^{(\mathrm{TN})}$
on $\mathit{\Delta}\varepsilon_p(0d_{3/2}\,\mbox{-}\,0d_{5/2})$
are investigated by plotting
$\mathit{\Delta}\varepsilon_p-\mathit{\Delta}\varepsilon_p^{(\mathrm{OPEP})}$
(the thin red dotted line)
and $\mathit{\Delta}\varepsilon_p-\mathit{\Delta}\varepsilon_p^{(\mathrm{OPEP})}
-\mathit{\Delta}\varepsilon_p^{(\mathrm{TN})}$ (the thin red dot-dashed line)
in Fig.~\ref{fig:dep_Z14}.
After contributions of $v_\mathrm{OPEP}^{(\mathrm{C})}$ and $v^{(\mathrm{TN})}$
are subtracted,
$\mathit{\Delta}\varepsilon_p(0d_{3/2}\,\mbox{-}\,0d_{5/2})$ with M3Y-P6
does not have strong $N$-dependence in $20\leq N\leq 28$,
becoming almost parallel to the corresponding D1M result.
Such similarity in the slope has been seen in the Ca isotopes
in Ref.~\cite{ref:NSM13}.
We confirm that the strong $N$-dependence
of $\mathit{\Delta}\varepsilon_p(0d_{3/2}\,\mbox{-}\,0d_{5/2})$
from $N=20$ to $28$ predominantly originates in $v^{(\mathrm{TN})}$,
aided by subsidiary contribution of $v_\mathrm{OPEP}^{(\mathrm{C})}$.

\subsection{$Z=40$}\label{subsec:TN-OPEP_Z40}

\begin{figure}[!h]
\centerline{\includegraphics[scale=1.5]{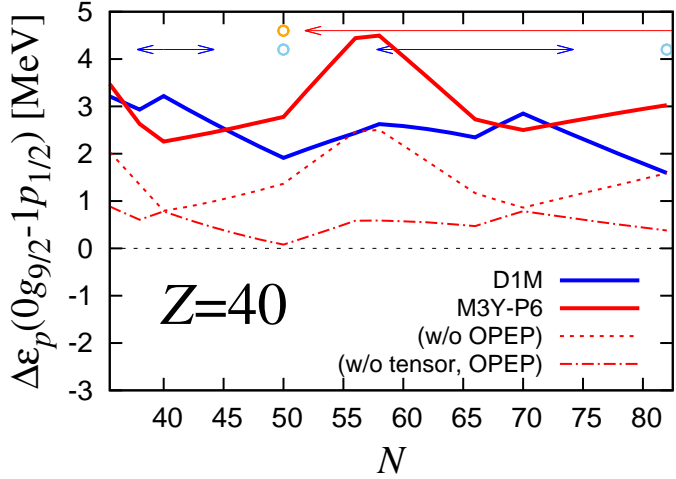}}
\caption{$\mathit{\Delta}\varepsilon_p(0g_{9/2}\,\mbox{-}\,1p_{1/2})$
 in the $Z=40$ nuclei.
 Blue and red solid lines are the results with D1M and M3Y-P6, respectively.
 For the latter, thin red dotted and dot-dashed lines represent
 effects of $v^{(\mathrm{TN})}$ and $v_\mathrm{OPEP}^{(\mathrm{C})}$
 as in Fig.~\protect\ref{fig:den_N56}.
 The region where $Z=40$ is magic in Fig.~\protect\ref{fig:magic_D1M}
 (Fig.~\protect\ref{fig:magic_M3Yp6}) is shown by the blue (red) arrows,
 and submagic
 by the skyblue (orange) circles at the top part of the figure.}
\label{fig:dep_Z40}
\end{figure}

The s.p. energy difference relevant to the $Z=40$ magicity is
$\mathit{\Delta}\varepsilon_p(0g_{9/2}\,\mbox{-}\,1p_{1/2})$.
We compare $N$-dependence of
$\mathit{\Delta}\varepsilon_p(0g_{9/2}\,\mbox{-}\,1p_{1/2})$
between D1M and M3Y-P6 in Fig.~\ref{fig:dep_Z40}.

We find opposite trends between the D1M and the M3Y-P6 results
on $\mathit{\Delta}\varepsilon_p(0g_{9/2}\,\mbox{-}\,1p_{1/2})$
in $40\leq N\leq 50$ as $n0g_{9/2}$ is occupied,
and in $70\leq N\leq 82$ as $n0h_{11/2}$ is occupied.
In addition, the rising tendency of
$\mathit{\Delta}\varepsilon_p(0g_{9/2}\,\mbox{-}\,1p_{1/2})$
at $N\approx 60$ in the M3Y-P6 result is not conspicuous with D1M.
It is notable that, after contributions of $v_\mathrm{OPEP}^{(\mathrm{C})}$
and $v^{(\mathrm{TN})}$ are subtracted,
the M3Y-P6 result becomes almost parallel to that of D1M.
On the contrary, once $v^{(\mathrm{TN})}$ is set in,
the s.p. energy difference has quite similar $N$-dependence
to the full result.
This clarifies significance of $v^{(\mathrm{TN})}$ in the $N$-dependence
of the shell gap at $Z=40$.
Relatively small $\mathit{\Delta}\varepsilon_p(0g_{9/2}\,\mbox{-}\,1p_{1/2})$
at $N=40$ contributes to the loss of the $Z=40$ magicity around $^{80}$Zr
viewed in Fig.~\ref{fig:magic_M3Yp6},
and relatively large $\mathit{\Delta}\varepsilon_p(0g_{9/2}\,\mbox{-}\,1p_{1/2})$
at $N=82$ to the persistence of the magicity around $^{122}$Zr.
The large $\mathit{\Delta}\varepsilon_p(0g_{9/2}\,\mbox{-}\,1p_{1/2})$
at $N\approx 60$
prevents the $Z=40$ magicity from being broken within the spherical HFB.
It will be an interesting future subject
whether and how the observed deformation at $N\gtrsim 60$ is accounted for,
under the sizable shell gap brought by $v^{(\mathrm{TN})}$.

\subsection{$Z=58$}\label{subsec:TN-OPEP_Z58}

\begin{figure}[!h]
\centerline{\includegraphics[scale=1.5]{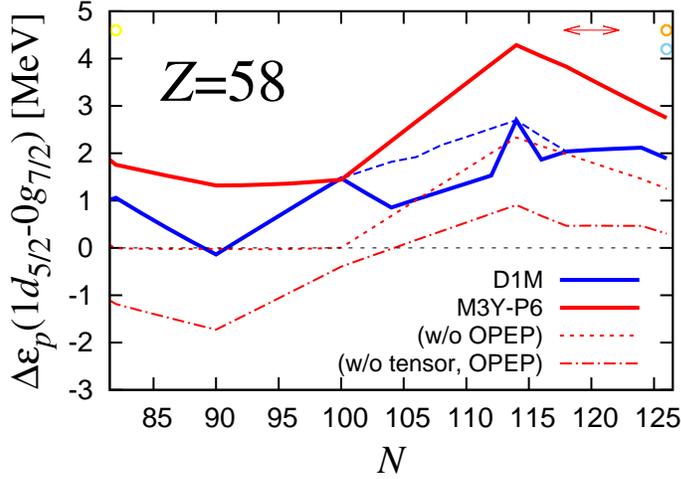}}
\caption{$\mathit{\Delta}\varepsilon_p(1d_{5/2}\,\mbox{-}\,0g_{7/2})$
 in the $Z=58$ nuclei.
 The blue dashed line is obtained with D1M by filling the s.p. levels
 in the same order as in the M3Y-P6 result.
 The red arrow, orange and yellow circles at the top of the figure
 correspond to the magic and submagic numbers
 in Fig.~\protect\ref{fig:magic_M3Yp6},
 while the skyblue circle to the submagic number
 in Fig.~\protect\ref{fig:magic_D1M}.
 See Fig.~\protect\ref{fig:den_N56} for other conventions.}
\label{fig:dep_Z58}
\end{figure}

The $Z=58$ magicity in the neutron-rich region of $N\gtrsim 110$
takes place because of the $p0g_{7/2}$ occupation,
to which the energy difference
$\mathit{\Delta}\varepsilon_p(1d_{5/2}\,\mbox{-}\,0g_{7/2})$ is relevant.
However, when comparing
$\mathit{\Delta}\varepsilon_p(1d_{5/2}\,\mbox{-}\,0g_{7/2})$
between D1M and M3Y-P6,
it should be noted that $n0i_{13/2}$ lies lower than $n2p_{3/2}$
in the spherical HF calculation with M3Y-P6 in $100\leq N\leq 120$,
while these two orbits are inverted with D1M except at $^{172}$Ce.
We therefore show $\mathit{\Delta}\varepsilon_p(1d_{5/2}\,\mbox{-}\,0g_{7/2})$
with D1M in which the s.p. levels were filled in the same ordering
as in the M3Y-P6 case,
by the blue dashed line in Fig.~\ref{fig:dep_Z58}.
If the neutron occupation is taken to be similar,
$\mathit{\Delta}\varepsilon_p(1d_{5/2}\,\mbox{-}\,0g_{7/2})$ with M3Y-P6
after removing the $v_\mathrm{OPEP}^{(\mathrm{C})}$ and $v^{(\mathrm{TN})}$ contributions
is almost parallel to $\mathit{\Delta}\varepsilon_p(1d_{5/2}\,\mbox{-}\,0g_{7/2})$
with D1M.
As seen in Fig.~\ref{fig:dep_Z58},
$v^{(\mathrm{TN})}$ gives rise to
large $\mathit{\Delta}\varepsilon_p(1d_{5/2}\,\mbox{-}\,0g_{7/2})$
at $N\approx 114$,
with a cooperative effect of $v_\mathrm{OPEP}^{(\mathrm{C})}$.
Although its degree depends on other channels of the interactions
as recognized by comparing Figs.~\ref{fig:magic_M3Yp6}
and \ref{fig:magic_M3Yp7},
it is expected that the $Z=58$ magicity is enhanced in $N\gtrsim 110$
because of $v^{(\mathrm{TN})}$ and $v_\mathrm{OPEP}^{(\mathrm{C})}$.

\subsection{Assessment of tensor force and OPEP effects}
\label{subsec:TN-OPEP_summary}

\begin{figure}[!h]
\centerline{\includegraphics[scale=0.6]{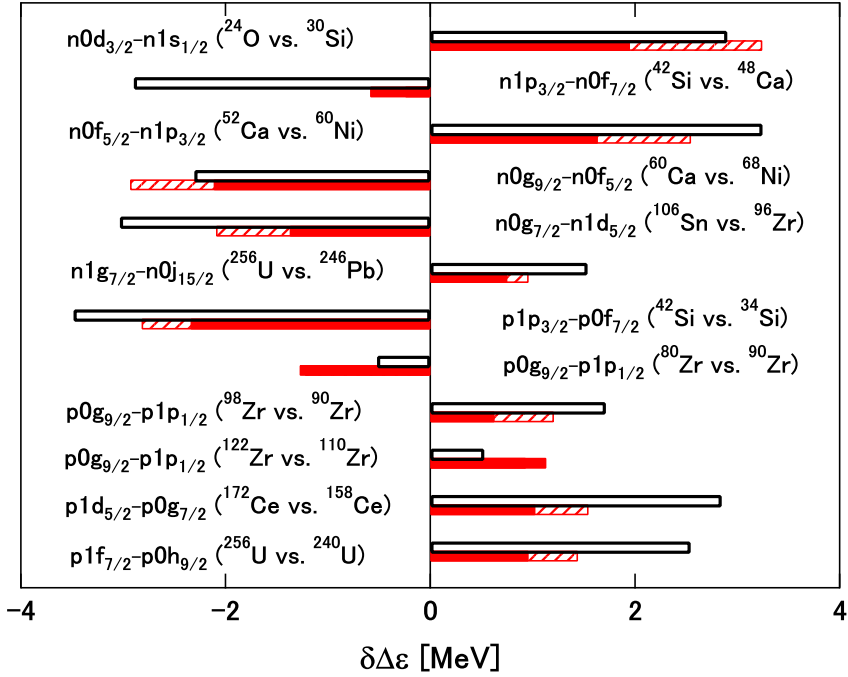}}
\caption{Difference of the shell gaps
 between two members of isotopes or isotones
 $\delta\mathit{\Delta}\varepsilon_{\tau_z}(j_2\,\mbox{-}\,j_1)$ (open bars),
 with contributions of $v^{(\mathrm{TN})}$ (red filled bars)
 and $v_\mathrm{OPEP}^{(\mathrm{C})}$ (red hatched bars) to it,
 obtained from the HF results with M3Y-P6.
 See text for details.}
\label{fig:ddspe_M3Yp6}
\end{figure}

To quantify the $Z$- or $N$-dependence of the shell gap,
we consider double difference of the s.p. energies
$\delta\mathit{\Delta}\varepsilon_{\tau_z}(j_2\,\mbox{-}\,j_1)$ ($\tau_z=p,n$):
$\mathit{\Delta}\varepsilon_{\tau_z}(j_2\,\mbox{-}\,j_1)$
at a certain nuclide $(Z_2,N_2)$
relative to that at a reference nuclide $(Z_1,N_1)$.
The reference nuclides are chosen so that they should be relatively close
to the $\beta$-stability line.
We can then view the effects of $v^{(\mathrm{TN})}$ and $v_\mathrm{OPEP}^{(\mathrm{C})}$
on $\delta\mathit{\Delta}\varepsilon_{\tau_z}(j_2\,\mbox{-}\,j_1)$
through the corresponding quantities
$\delta\mathit{\Delta}\varepsilon^{(\mathrm{TN})}_{\tau_z}(j_2\,\mbox{-}\,j_1)$
and $\delta\mathit{\Delta}\varepsilon^{(\mathrm{OPEP})}_{\tau_z}(j_2\,\mbox{-}\,j_1)$.
Figure~\ref{fig:ddspe_M3Yp6} summarizes the effects
of $v^{(\mathrm{TN})}$ and $v_\mathrm{OPEP}^{(\mathrm{C})}$
on $\delta\mathit{\Delta}\varepsilon_{\tau_z}(j_2\,\mbox{-}\,j_1)$
in the HF results with the M3Y-P6 interaction,
by selecting the nuclei at which $v^{(\mathrm{TN})}$ affects the magic numbers.
We find that
$\delta\mathit{\Delta}\varepsilon^{(\mathrm{OPEP})}_{\tau_z}(j_2\,\mbox{-}\,j_1)$
usually has equal sign but does not exceed 
$\delta\mathit{\Delta}\varepsilon^{(\mathrm{TN})}_{\tau_z}(j_2\,\mbox{-}\,j_1)$
as long as sizable,
although it should not be discarded in many cases.

We look at some of the individual results in Fig.~\ref{fig:ddspe_M3Yp6}
for illustration.
The $\mathit{\Delta}\varepsilon_n(0d_{3/2}\,\mbox{-}\,1s_{1/2})$ value at $^{24}$O
relative to that at $^{30}$Si is shown
at the top row of Fig.~\ref{fig:ddspe_M3Yp6}.
This seems to account for the $N=16$ magicity at $^{24}$O.
We obtain $\delta\mathit{\Delta}\varepsilon_n(0d_{3/2}\,\mbox{-}\,1s_{1/2})
=2.9\,\mathrm{MeV}$ with M3Y-P6,
which means that the $N=16$ shell gap is larger at $^{24}$O
than at $^{30}$Si by $2.9\,\mathrm{MeV}$.
Since $\delta\mathit{\Delta}\varepsilon^{(\mathrm{TN})}_n(0d_{3/2}\,\mbox{-}\,
1s_{1/2})=2.0\,\mathrm{MeV}$
and $\delta\mathit{\Delta}\varepsilon^{(\mathrm{OPEP})}_n(0d_{3/2}\,\mbox{-}\,1s_{1/2})
=1.3\,\mathrm{MeV}$,
it is difficult to obtain the enhancement of the shell gap
without $v^{(\mathrm{TN})}$ and $v_\mathrm{OPEP}^{(\mathrm{C})}$.
The gap could even be reduced from $^{30}$Si to $^{24}$O without them.

At the second top row of Fig.~\ref{fig:ddspe_M3Yp6},
$\delta\mathit{\Delta}\varepsilon_n(1p_{3/2}\,\mbox{-}\,0f_{7/2})$
obtained as $\mathit{\Delta}\varepsilon_n(1p_{3/2}\,\mbox{-}\,0f_{7/2})$
at $^{42}$Si relative to that at $^{48}$Ca is presented.
The negative value of
$\delta\mathit{\Delta}\varepsilon_n(1p_{3/2}\,\mbox{-}\,0f_{7/2})$
indicates quenching of the $N=28$ shell gap at $^{42}$Si.
However, $v^{(\mathrm{TN})}$ gives only $-0.6\,\mathrm{MeV}$
to the full $\delta\mathit{\Delta}\varepsilon_n(1p_{3/2}\,\mbox{-}\,0f_{7/2})
\,(=-2.9\,\mathrm{MeV})$,
and $v_\mathrm{OPEP}^{(\mathrm{C})}$ contribution is small but positive
(not visible in Fig.~\ref{fig:ddspe_M3Yp6}).
Though the $v^{(\mathrm{TN})}$ effect should not be ignored
in describing the $N=28$ shell erosion,
it is not necessarily dominant.

We next discuss $\delta\mathit{\Delta}\varepsilon_p(0g_{9/2}\,\mbox{-}\,1p_{1/2})$
evaluated for $^{80}$Zr relative to $^{90}$Zr.
We have $-0.5\,\mathrm{MeV}$ in the full M3Y-P6 result,
while $\delta\mathit{\Delta}\varepsilon^{(\mathrm{TN})}_p(0g_{9/2}\,\mbox{-}\,1p_{1/2})
=-1.3\,\mathrm{MeV}$.
Namely,
the sign of $\delta\mathit{\Delta}\varepsilon_p(0g_{9/2}\,\mbox{-}\,1p_{1/2})$
is inverted owing to $v^{(\mathrm{TN})}$,
which could be crucial in the erosion of the $Z=40$ magicity at $N\approx 40$.
Sign inversion due to $v^{(\mathrm{TN})}$ is found
also in the comparison between $^{122}$Zr and $^{110}$Zr,
which could be important to persistence of the $N=82$ magicity
around $^{122}$Zr.

Several points on the tensor-force and OPEP effects are confirmed
from Fig.~\ref{fig:ddspe_M3Yp6}:
\begin{itemize}
\item The tensor force often plays a significant role
 in the $Z$- or $N$-dependence of the shell gap,
 accounting for appearance and disappearance of magicity.
\item The central spin-isospin channel from the OPEP
 tends to enhance the tensor-force effect.
 Strong $Z$- or $N$-dependence of the shell gap often coincides
 with their cooperative contribution.
\item These effects strongly appear as an orbit with relatively high $\ell$
 is occupied.
\end{itemize}

\section{Conclusion}\label{sec:conclusion}

We predict magic and submagic numbers in wide range of the nuclear chart
from the spherical mean-field calculations
with the M3Y-P6 and P7 semi-realistic $NN$ interactions,
which contain the tensor force from the $G$-matrix
as well as the central spin-isospin channel from the OPEP.
The magic numbers are identified by vanishing pair correlations
in the HFB results,
and the submagic numbers by small difference
between the HFB and the HF energies.
Although deformation degrees of freedom are not explicitly taken into account,
the semi-realistic interactions describe
the erosion of the $N=20$ and $28$ magicity in the proton-deficient region,
as well as the emergence of the $N=16$ and $32$ magicity.
In addition to the known magic numbers,
possible magicity at $N=40$, $56$, $90$, $124$, $172$, $178$, $164$, $184$
and $Z=14$, $16$, $34$, $38$, $40$, $58$, $64$, $92$, $120$, $124$, $126$
has been argued,
in comparison with similar prediction obtained
with the Gogny D1S and D1M interactions.
The predictions with M3Y-P6 do not contradict to the known data apparently,
except at $^{32}$Mg and the Zr isotopes with $60\leq N\lesssim 70$.
Although calculations including the deformation degrees of freedom
are needed for complete understanding of the magic numbers,
this work will be useful in selecting candidates of the magic numbers
and in overviewing how the magic numbers can distribute over the nuclear chart.
 
By analyzing the shell gaps,
roles of the tensor force and of the central spin-isospin channel from the OPEP
are investigated.
It is confirmed that the tensor force often plays a significant role
in the $Z$- or $N$-dependence of the shall gap,
accounting for appearance and disappearance of magicity,
and the central spin-isospin channel tends to enhance the tensor-force effect
for emergence of the magicity.
The present results are qualitatively consistent
with Refs.~\cite{ref:Vtn,ref:Vst},
although quantitative aspects should not be underestimated
because they make difference in certain cases.

It will be interesting to study effects of deformation on the magicity
with the semi-realistic interactions,
and to investigate whether and how the discrepancy
between the current results and the data in several nuclei
could be resolved.

\section*{Acknowledgment}

The authors are grateful to T.~Inakura
for discussions and his assistance in drawing the figures.
This work is financially supported
as Grant-in-Aid for Scientific Research (C), Nos.~22540266 and 25400245,
by Japan Society for the Promotion of Science,
and as Grant-in-Aid for Scientific Research on Innovative Areas,
No.~24105008, by The Ministry of Education, Culture, Sports, Science
and Technology, Japan.
Numerical calculations are performed on HITAC SR16000s
at Institute of Management and Information Technologies in Chiba University,
at Yukawa Institute for Theoretical Physics in Kyoto University,
at Information Technology Center in University of Tokyo,
and at Information Initiative Center in Hokkaido University.


%

\vfill\pagebreak

\end{document}